\documentclass[
final
 , nomarks
]{dmtcs-episciences}

\usepackage[utf8]{inputenc}
\usepackage{subfigure}

\usepackage{enumerate}
\usepackage{scalefnt}
\newcommand\smaller[2][0.85]{{\scalefont{#1}#2}}
\newtheorem{theorem}{Theorem}

\usepackage[round]{natbib}

\author{Timothy Highley
  \and Hoang Le}
\title[Tropical Vertex-Disjoint Cycles of a Vertex-Colored Digraph]{Tropical Vertex-Disjoint Cycles of a Vertex-Colored Digraph: Barter Exchange with Multiple Items Per Agent}
\affiliation{
  La Salle University, Philadelphia, Pennsylvania, USA}
\received{2017-3-17}
\revised{2017-8-3, 2018-4-6}
\accepted{2018-6-26}
\begin{document}
\publicationdetails{20}{2018}{2}{1}{3186}	
\maketitle
\begin{abstract}
  In a barter exchange market, agents bring items and seek to exchange their items with one another. Agents may agree to a $k$-way exchange involving a cycle of $k$ agents.  A barter exchange market can be represented by a digraph where the vertices represent items and the edges out of a vertex indicate the items that an agent is willing to accept in exchange for that item. It is known that the problem of finding a set of vertex-disjoint cycles with the maximum total number of vertices (MAX-SIZE-EXCHANGE) can be solved in polynomial time. We consider a barter exchange where each agent may bring multiple items, and items of the same agent are represented by vertices with the same color.  A set of cycles is said to be tropical if for every color there is a cycle that contains a vertex of that color.  We show that the problem of determining whether there exists a tropical set of vertex-disjoint cycles in a digraph (TROPICAL-EXCHANGE) is NP-complete and APX-hard. This is equivalent to determining whether it is possible to arrange an exchange of items among agents such that every agent trades away at least one item. TROPICAL-MAX-SIZE-EXCHANGE is a similar problem, where the goal is to find a set of vertex-disjoint cycles that contains the maximum number of vertices and also contains all of the colors in the graph. We show that this problem is likewise NP-complete and APX-hard. For the restricted case where there are at most two vertices of each color (corresponding to a restriction that each agent may bring at most two items), both problems remain NP-hard but are in APX. Finally, we consider MAX-SIZE-TROPICAL-EXCHANGE, where the set of cycles must primarily include as many colors as possible and secondarily include as many vertices as possible. We show that this problem is NP-hard.
\end{abstract}

\section{Introduction}
\label{se:intro}

Consider a barter exchange where many different agents each bring multiple items to trade.  For each item at the barter exchange, the agent lists which other items may be accepted in trade for that item.  Based on the lists, an algorithm determines which items are traded for which other items.  Instead of restricting exchanges to one-to-one trades, $k$-way trades are also permitted, where $k$ is limited only by the number of items at the swap meet.  If $k$-way trades are permitted, the traders potentially give items to and receive items from different agents.

This barter exchange can be represented as a directed graph where each item brought to the exchange is a vertex and there are edges from each item to those items that may be accepted in trade for that item.  The problem of determining the maximum possible number of items that can be simultaneously exchanged is equivalent to finding a set of vertex-disjoint cycles that maximizes the total number of vertices in the cycles.  (Throughout this paper, all cycles are assumed to be simple cycles.) 

This problem is equivalent to the Assignment Problem, which can be solved in polynomial time. It has also been called \smaller{MAX-SIZE-EXCHANGE} ~\citep*{biro2009maximum}.  The decision and optimization versions of the problem are defined as follows. 
\begin{description}
	\item[Problem 1a.] \smaller{EXCHANGE-x}
	\item[Input:] A directed graph $G$ and an integer $x$
	\item[Question:] Is there a set of vertex-disjoint cycles that includes at least $x$ vertices?
\end{description}

\begin{description}
	\item[Problem 1b.] \smaller{MAX-SIZE-EXCHANGE}
	\item[Input:] A directed graph $G$
	\item[Output:] A set of vertex-disjoint cycles that includes as many vertices as possible
\end{description}

If there are many items, it is very likely that there are multiple solutions that tie for the maximum number of items traded.  How should the tie be broken?  As a secondary criterion, it may be desirable to choose, from among those item-maximizing solutions, a solution that maximizes the number of agents who trade away at least one item.  This can be accomplished by converting the previously described directed graph into a vertex-colored digraph, with each participant having a unique color.  The problem, then, is equivalent to finding a set of vertex-disjoint cycles with the maximum total number of vertices in the cycles, and in the case of ties, finding a solution that secondarily contains the maximum number of colors. A subset of vertices from a vertex-colored graph is called \textit{tropical} if it includes every color in the vertex-colored graph ~\citep{ang16domset}.  We define the decision and optimization versions of the \smaller{TROPICAL-MAX-SIZE-EXCHANGE} as follows.

\begin{description}
	\item[Problem 2a.] \smaller{TROPICAL-MAX-SIZE-EXCHANGE-d} (\smaller{TMaxEx-d})
	\item[Input:] A vertex-colored digraph $G$
	\item[Question:] Is there a set of vertex-disjoint cycles $S$ such that no other set of vertex-disjoint cycles contains more vertices in the cycles, and for every color in the graph there is a cycle in $S$ that contains a vertex of that color? 
\end{description}

\begin{description}
	\item[Problem 2b.] \smaller{TROPICAL-MAX-SIZE-EXCHANGE-o} (\smaller{TMaxEx-o})
	\item[Input:] A vertex-colored digraph $G$
	\item[Output:] A set of vertex-disjoint cycles that includes as many colors as possible, subject to the restriction that there does not exist another set of vertex-disjoint cycles that has more vertices
\end{description}

If the goal is instead to simply maximize the number of agents who trade away at least one item (with no regard for the total number of items traded away), that is equivalent to the problem of finding a set of vertex-disjoint cycles that collectively contain the maximum number of colors. To this end, we define the decision and optimization versions of \smaller{TROPICAL-EXCHANGE} as follows.  

\begin{description}
	\item[Problem 3a.] \smaller{TROPICAL-EXCHANGE-d (TEx-d)}
	\item[Input:] A vertex-colored digraph $G$
	\item[Question:] Is there a set of vertex-disjoint cycles such that for every color there is a cycle that contains a vertex of that color?
\end{description}

\begin{description}
	\item[Problem 3b.] \smaller{TROPICAL-EXCHANGE-o (TEx-o)}
	\item[Input:] A vertex-colored digraph $G$.
	\item[Output:] A set of vertex-disjoint cycles such that the total number of vertex colors in the cycles is maximized
\end{description}

We show that both \smaller{TMaxEx-d} and \smaller{TEx-d} are NP-complete via a reduction from \smaller{CNFSAT}. We show that both \smaller{TMaxEx-o} and \smaller{TEx-o} are APX-hard. 

We also consider restricted cases where each agent is permitted to bring at most $j$ items to the barter exchange.  We then maximize the number of agents who get to trade an item:

\begin{description}
	\item[Problem 4.] \smaller{j-PER-COLOR-TROPICAL-EXCHANGE (jPC-TEx)}
	\item[Input:] A vertex-colored digraph $G$ with at most $j$ vertices of each color
	\item[Output:] A set of vertex-disjoint cycles such that the total number of vertex colors in the cycles is maximized
\end{description}

\begin{description}
	\item[Problem 5.] \smaller{j-PER-COLOR-TROPICAL-MAX-SIZE-EXCHANGE (jPC-TMaxEx)}
	\item[Input:] A vertex-colored digraph $G$ with at most $j$ vertices of each color
	\item[Output:] A set of vertex-disjoint cycles that includes as many colors as possible, subject to the restriction that no other set of vertex-disjoint cycles has more vertices
\end{description}

These problems can be trivially solved in polynomial time if $j=1$, but even if $j=2$, a reduction from \smaller{MAX-2-SAT} shows that these problems are NP-hard. They are, however, in APX for any fixed $j$. 

Finally, we define \smaller{MAX-SIZE-TROPICAL-EXCHANGE}, which reverses the criteria of \smaller{TMaxEx-o}. In \smaller{TMaxEx-o}, the first criterion is maximizing the total number of vertices in the cycles, and the second criterion is maximizing the total number of colors.  \smaller{MAX-SIZE-TROPICAL-EXCHANGE} first maximizes the total number of colors in the cycles and secondarily maximizes the total number of vertices.  We show that it is NP-hard and APX-hard. We define the decision and optimization versions of \smaller{MAX-SIZE-TROPICAL-EXCHANGE} as follows.

\begin{samepage}
\begin{description}
	\item[Problem 6a.] \smaller{MAX-SIZE-TROPICAL-EXCHANGE-d} (\smaller{MaxTEx-d})
	\item[Input:] A vertex-colored digraph $G$ and an integer $x$.
	\item[Question:] Is there a set of vertex-disjoint cycles $S$ with a total of at least $x$ vertices in the cycles such that no other set of vertex-disjoint cycles contains more colors in the cycles? 
\end{description}
\end{samepage}
\begin{description}
	\item[Problem 6b.] \smaller{MAX-SIZE-TROPICAL-EXCHANGE-o} (\smaller{MaxTEx-o})
	\item[Input:] A vertex-colored digraph $G$.
	\item[Output:] A set of vertex-disjoint cycles $S$ such that no other set of vertex-disjoint cycles contains more colors, and no other set with the same number of colors contains more vertices.
\end{description}

Barter exchanges as described in this paper regularly occur online. For example, since 2006 an average of 10 barter exchanges per month have been organized at boardgamegeek.com, with a total of over 500,000 items offered for trade. When participants are informed of the results, they exchange items either in person or by mailing items to each other.  Usually, a participant will give an item to one person and receive an item from a different person.  Maximizing the number of items traded is the primary goal, and is accomplished using a solution to the Assignment Problem.  Maximizing the number of users that trade at least one item is the secondary goal, and this is addressed by solving the Assignment Problem multiple times and after a set number of iterations selecting the solution that yields the greatest number of users trading.  However, there is no guarantee that the chosen solution actually maximizes the number of users who trade at least one item.  This paper demonstrates that, in general, the problem of maximizing the number of users trading cannot be solved in polynomial time unless $P=NP$. Other websites, such as barterquest.com, swapdom.com, netcycle.com, and rehash.com, have offered similar services in a more general context.

\section{Related Work}\label{se:relatedwork}
The Assignment Problem is the minimum weight bipartite perfect matching problem.  It is a well-known fundamental optimization problem that can be solved in polynomial time.  The polynomial bound was first shown by Jacobi ~\citep[see][]{jac07}, but was not widely known until Kuhn's publication of the Hungarian Algorithm ~\citep{kuhn55}.  More efficient algorithms for the Assignment Problem have been discovered since then ~\citep{daitch08}~\citep*{lee14}.  It is known that \smaller{MAX-SIZE-EXCHANGE} can be solved by reducing it to the Assignment Problem ~\citep{abraham2007clearing}~\citep{biro2009maximum}.

\smaller{MAX-SIZE-EXCHANGE} and related problems have been studied extensively in the context of kidney exchange programs.  In kidney exchange programs, it is impractical to allow cycles of unbounded length because it is important for all surgeries to happen at the same time.  As a result, most work related to kidney exchanges addresses the \smaller{MAX SIZE $\leq k$-WAY EXCHANGE} problem, which is APX-complete when $k \geq 3$ ~\citep{abraham2007clearing}~\citep{biro2009maximum}.  Work on the kidney exchange problem includes investigation of approximability ~\citep*{biro2007inapproximability}, techniques to mitigate failure after matches have been determined ~\citep*{dickerson2013failure}, dynamic exchanges ~\citep*{unver2010dynamic}, and incentive-compatible mechanisms to ensure that hospitals publish all incompatible pairs ~\citep{ashlagi2015mix}.  The problems discussed in this paper differ from the kidney exchange problem because we are not concerned with the length of the cycles, and because of the introduction of colored vertices. 

We explore problems that are similar to the Assignment Problem and kidney exchange problem but based on vertex-colored graphs. Vertex-colored graphs have also been called tropical graphs ~\citep{foucaud2016tropical}.  As mentioned earlier, a subset of vertices from a vertex-colored graph is called tropical if it includes every color in the vertex-colored graph.  Tropical connected subgraphs, tropical dominating sets, and tropical homomorphisms have been studied ~\citep{ang16domset}~\citep{ang16subgraph}~\citep{foucaud2016tropical}.  In this paper, we are concerned with sets of vertex-disjoint cycles.  We say that a set of cycles is tropical if every color in the vertex-colored graph is represented by a vertex in at least one cycle.

We study the total number of vertices and colors in a set of vertex-disjoint cycles, but not the total number of cycles in the set.  Cycle packing is concerned with the total number of cycles, and it has been studied extensively for both edge-disjoint and vertex-disjoint cycles ~\citep{bodlaender2009kernel}~\citep{krivelevich2007approximation}~\citep*{pedroso2014maximizing}. 

There is other work on graphs and colors that is not directly related to our work, but we provide representative citations to help readers delineate the differences. For instance, ~\citet{fellows2011upper} and ~\citet*{dondi2011complexity} explored pattern-matching in vertex-colored graphs. ~\citet{coudert2007shared} explored complexity and approximability properties of edge-colored graphs. Our work assumes that the color of a vertex is an inherent property of the graph.  Graph-coloring is a separate topic where colors are added to a graph such that adjacent nodes have different colors. The Four-Color Theorem is a famous result from graph-coloring ~\citep*{appel1977every1}~\citep*{appel1977every2}.  Coloring graphs are graphs where each node represents a possible coloring of another graph ~\citep*{beier16classifying}.  More recent graph-coloring work includes the search for rainbow connections ~\citep{chartrand2008rainbow}~\citep*{li2012rainbow}~\citep*{li2013rainbow}.  Colorful paths are paths that are both rainbow paths (all vertices in the path have different colors) and tropical paths (all colors in the graph are represented in the path) ~\citep{akbari2011colorful}.

\section{\smaller{CNFSAT} as a Vertex-Colored Graph}\label{se:rep_as_graph}
Consider an instance of \smaller{CNFSAT} with $q$ clauses.  Let $x_1$, ..., $x_n$ be the $n$ variables in the expression.  For each variable, create a vertex.  Give all of these vertices the same color, and refer to them as the``variable vertices.''  On each vertex, create two edges that loop back to the same vertex.  Label one edge TRUE and one edge FALSE on each vertex.  

Assign each clause of the \smaller{CNFSAT} instance a separate color: $c_1$, $c_2$, ..., $c_q$.  For each literal in the clause, create a vertex that has the clause's color.  Refer to these vertices as ``literal vertices."  For negative literals, put the corresponding vertex on the FALSE loop of the variable vertex for that variable by removing one edge from that loop and adding the new literal vertex along with two edges so that the FALSE loop remains a single cycle from the variable vertex back to itself.  Similarly, for positive literals put the literal vertex on the TRUE loop of the variable vertex for that variable.  

\begin{figure}[htbp]
	\begin{center}
		\includegraphics[width=12cm]{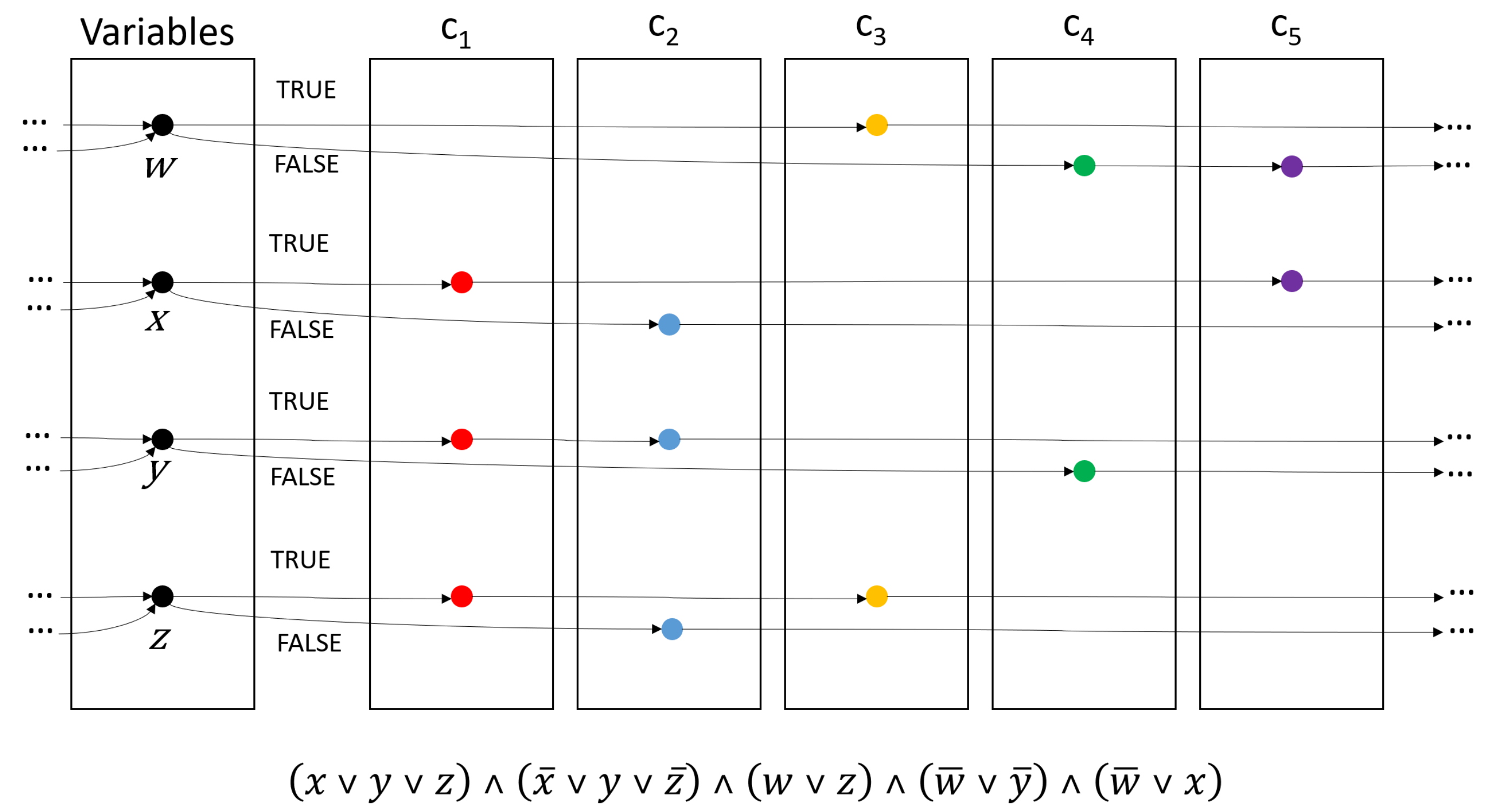}
	\end{center}
	\caption{CNFSAT instance represented as a vertex-colored directed graph.}
	\label{fig:CNF_as_digraph}	
\end{figure}

Figure \ref{fig:CNF_as_digraph} depicts an example graph for a given \smaller{CNFSAT} instance.

\section{\smaller{CNFSAT} reduces to \smaller{TROPICAL-EXCHANGE-d}}\label{se:reduce_TROPICAL-EXCHANGE}
Constructing a graph for a \smaller{CNFSAT} instance can clearly be accomplished in polynomial time.  Each variable only adds one vertex and two edges to the graph, and each clause only adds a vertex and an edge for each literal in the clause.  

Based on the way that the graph is constructed, there can be at most $n$ vertex-disjoint cycles.  Each variable vertex is part of two cycles (the TRUE loop and the FALSE loop), so a set of vertex-disjoint cycles can only contain one or the other.  There are no other cycles in the graph.  The cycle that is chosen for the set of vertex-disjoint cycles corresponds to the truth value of that variable in the expression.  If we have a polynomial time algorithm for \smaller{TROPICAL-EXCHANGE-d} (\smaller{TEx-d}), then solving \smaller{TEx-d} will also solve \smaller{CNFSAT} in polynomial time.  This is because each clause's color will be in a cycle if and only if a variable is set to a truth value that satisfies that clause.  If all colors can be in the set of vertex-disjoint cycles, then all clauses and thus the entire expression is satisfiable.  Since \smaller{CNFSAT} is NP-complete, \smaller{TEx-d} is NP-hard. \smaller{TEx-d} is also in NP (it is trivial to verify that all colors are in the solution), so \smaller{TEx-d} is NP-complete.

\begin{theorem}
	\smaller{TEx-d} is NP-complete.
\end{theorem}

\section{\smaller{CNFSAT} reduces to  \smaller{TROPICAL-MAX-SIZE-EXCHANGE}}\label{se:reduce_TROPMAX-EXCHANGE}
\smaller{EXCHANGE-x} can be solved in polynomial time, but \smaller{TROPICAL-EXCHANGE-d} is NP-complete.  What about \smaller{TROPICAL-MAX-SIZE-EXCHANGE-d (TMaxEx-d)}?  That is, if we restrict the search space to only those sets of cycles that maximize the total number of vertices in the cycles, then does that make it easier to determine whether it is possible for a set of vertex-disjoint cycles to cover all of the colors in the graph? 

The previous reduction does not work for \smaller{TMaxEx-d} due to the fact that solving \smaller{TMaxEx-d} for the constructed graph does not solve \smaller{CNFSAT}.  Because \smaller{TMaxEx-d} first maximizes the total number of vertices in the cycles, it may be the case that the \smaller{CNFSAT} instance is satisfiable even though the solution to \smaller{TMaxEx-d} for the graph does not include all of the colors.  That would be the case if the only way to include all of the colors in the graph in the cycles is to accept a lower total number of vertices in the cycles.

We can adjust the graph by adding ``balance vertices'' that all have the same color (the ``balance color'').  First, we identify the cycle with the most vertices.  In the case of a tie, we choose any of the tied cycles.  We add one balance vertex to this cycle and then add balance vertices to all other cycles so that every cycle has the same number of vertices.  These balance vertices accomplish two tasks:

\begin{itemize}
	\item The number of vertices in each cycle has no impact on the solution produced by \smaller{TMaxEx-d}.  With the addition of the balance vertices, every cycle has the same number of vertices, so every possible combination of vertex-disjoint cycles has the same total number of vertices, assuming one cycle per variable vertex.  As a result, the secondary criterion (number of colors in the set of cycles) determines the solution.
	\item The balance color has no impact.  Because every cycle has a balance vertex, the balance color is guaranteed to be in the set of vertex-disjoint cycles, and the balance color will not have an impact on the final solution.
\end{itemize}

If all possible combinations of vertex-disjoint cycles have the same number of total vertices, then by the same logic described for the previous reduction, the solution to \smaller{TMaxEx-d} for the constructed graph will also solve \smaller{CNFSAT}.  

\begin{figure}[htbp]
	\begin{center}
		\includegraphics[width=12cm]{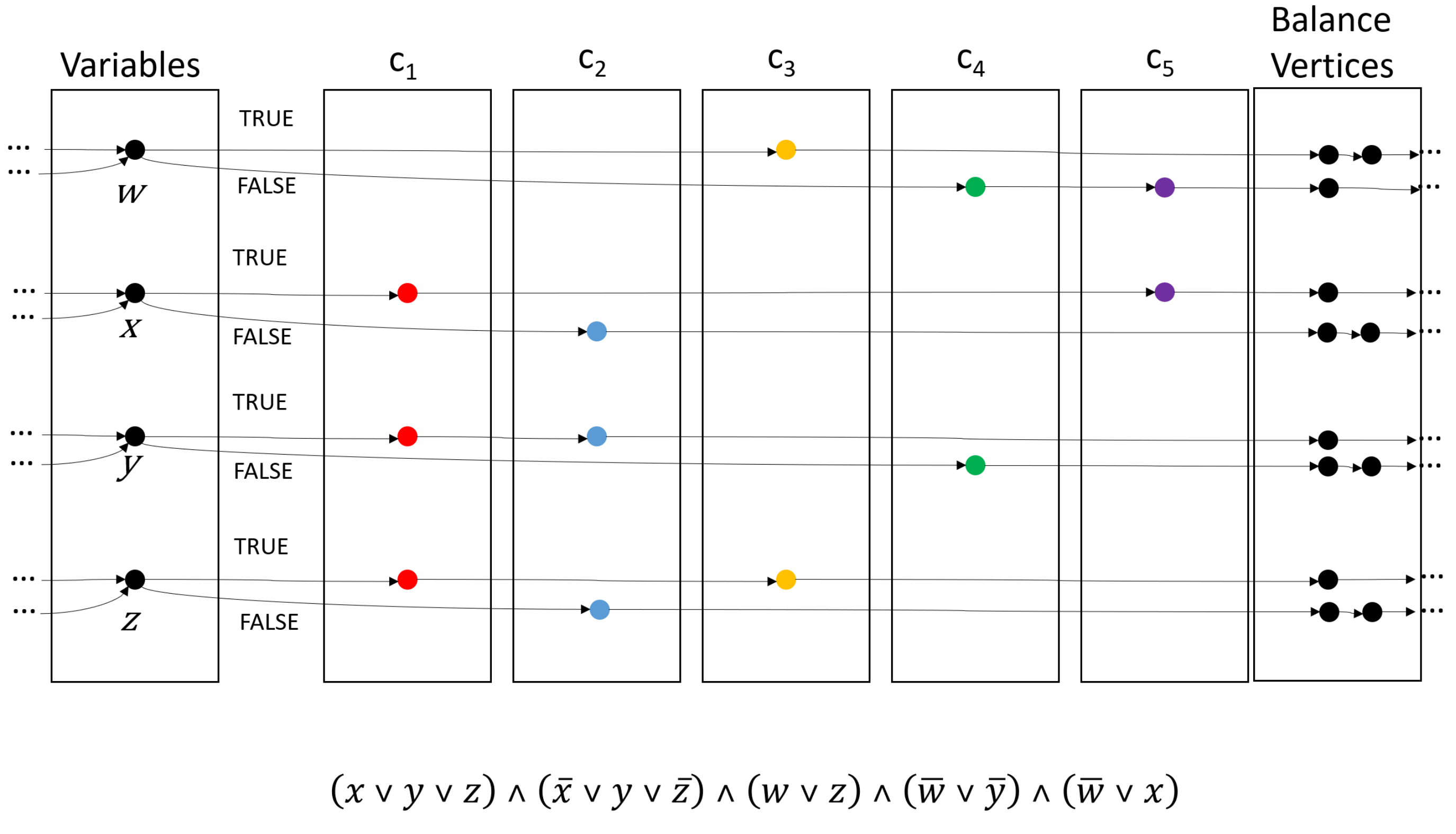}
	\end{center}
	\caption{\smaller{CNFSAT} graph with the addition of Balance Vertices.}
	\label{fig:CNF_with_balance}
\end{figure}

Figure \ref{fig:CNF_with_balance} depicts an example graph for a given \smaller{CNFSAT} instance.

\smaller{TMaxEx-d} is easily shown to be in NP.  A proposed solution to an instance of \smaller{TMaxEx-d} can be verified by first solving \smaller{MAX-SIZE-EXCHANGE-o} for the same graph, which can be accomplished in polynomial time.  The next step is verifying that the proposed solution to \smaller{TMaxEx-d} and the solution produced by \smaller{MAX-SIZE-EXCHANGE-o} have the same number of vertices in their respective sets of vertex-disjoint cycles. Assuming they do, it is simple to verify that all colors appear in the proposed solution.  Therefore, \smaller{TMaxEx-d} is NP-complete.

\begin{theorem}
	\smaller{TMaxEx-d} is NP-complete.
\end{theorem}

\section{\smaller{TROPICAL-EXCHANGE} and \smaller{TROPICAL-MAX-SIZE-EXCHANGE} are APX-hard}

\smaller{MAX-3-SAT3} is a well-known APX-complete problem that restricts \smaller{CNFSAT} to at most three literals per clause, and each literal can appear in only three clauses.  Additionally, it is an optimization problem instead of a decision problem.  Section \ref{se:reduce_TROPICAL-EXCHANGE} contains a reduction from \smaller{CNFSAT} to \smaller{TROPICAL-EXCHANGE}.  A reduction from \smaller{MAX-3-SAT3} to \smaller{TEx-o} works like the reduction in Section \ref{se:reduce_TROPICAL-EXCHANGE}.  We show that such a reduction is an L-reduction, so we can conclude that \smaller{TEx-o} is APX-hard.

Some additional definitions from \cite{crescenzi1997short} and \cite{papadimitrou1991optimization} are needed.  A reduction from problem A to problem B consists of a function $f$ that maps an instance of problem A to an instance of problem B, and a function $g$ that maps a solution of problem B to a solution of problem A.  The set of instances of problem A is $I_A$.  The set of feasible solutions of an instance $x \in I_A$ is $sol(x)$.  The measure of the quality of a solution $y \in sol(x)$ of an instance $x \in I_A$ is denoted by $m(x, y)$.  If a solution $y$ is optimal, then $m(x,y) = opt(x)$.  The absolute error $E(x,y) = |opt(x) - m(x,y)|$. 

A reduction is said to be an L-reduction if it meets the following criteria.

\begin{enumerate}
	\item For any instance of A and corresponding instance of B, the optimal solution of B is within a constant factor of the optimal solution of A.  More formally:
	\begin{math}
	\forall x \in I_A, opt_B (f(x)) \le \alpha \cdot opt_A (x)
	\end{math}
	\item For any $x$ that is an instance of problem A and any $y$ that is the solution to corresponding instance of problem B, the absolute error of the proposed solution to A is within a constant factor of the absolute error of y.  More formally:
	\begin{math}
	\forall x \in I_A \forall y \in sol_B(f(x)), E_A(x, g(x,y)) \le \beta E_B(f(x), y)
	\end{math}
\end{enumerate}

For our purposes, problem A is \smaller{MAX-3-SAT3} and problem B is \smaller{TEx-o}.  For \smaller{TEx-o} and \smaller{TMaxEx-o}, $m(x,y)$ measures the number of colors in the solution. The reduction in Section \ref{se:reduce_TROPICAL-EXCHANGE} satisfies criterion 1 because the number of colors in the graph is equal to the number of clauses in the expression plus one.  Similarly, the number of colors in the solution to \smaller{TEx-o} is equal to one plus the number of satisfied clauses in the solution to \smaller{MAX-3-SAT3}.  For each instance of \smaller{MAX-3-SAT3} it is necessary to find a value for $\alpha$ that satisfies criterion 1.  If, for example, $\alpha = 3$, then the criterion is satisfied for all instances.

The reduction satisfies criterion 2 because the absolute errors for the two problems are equal.  The number of clauses not satisfied is exactly equal to the number of colors that do not appear in the proposed solution to \smaller{TEx-o}.  For each instance it is necessary to find a value for $\beta$ that satisfies criterion 2.  If $\beta = 1$, criterion 2 is satisfied for all instances.

Therefore, we conclude that it is an L-reduction and \smaller{MAX-3-SAT3} $\le_L$ \smaller{TEx-o}. Because an L-reduction preserves membership in PTAS \citep[Proposition 7]{crescenzi1997short}, we can say that \smaller{TEx-o} is APX-hard and there is no PTAS for \smaller{TEx-o}.  A similar argument can be made about the reduction in Section \ref{se:reduce_TROPMAX-EXCHANGE}.  That allows us to conclude that \smaller{MAX-3-SAT3} $\le_L$ \smaller{TMaxEx-o}, and that \smaller{TMaxEx-o} is APX-hard and has no PTAS.

\section{\smaller{TROPICAL-EXCHANGE} with only 2 vertices per color}

We now consider the problems \smaller{jPC-TEx} and \smaller{jPC-TMaxEx}, which are special cases of \smaller{TROPICAL-EXCHANGE-o} and \smaller{TROPICAL-MAX-SIZE-EXCHANGE-o}, respectively. For these problems, there can be at most $j$ vertices of each color. In terms of a barter exchange, this corresponds to an organized exchange where each person is permitted to bring at most $j$ items.  We will show that these restricted cases are in APX and, unless $j=1$, they are NP-hard.

There is a trivial algorithm to show that both \smaller{2PC-TEx} and \smaller{2PC-TMaxEx} are in APX. Simply solve \smaller{MAX-SIZE-EXCHANGE} for the given graph, completely ignoring vertex colors. Because there are only two vertices in the graph for each color, the solution to \smaller{MAX-SIZE-EXCHANGE} is guaranteed to have at least half as many colors as the maximum number of colors possible in a solution to \smaller{2PC-TEx} or \smaller{2PC-TMaxEx}.  Therefore it is a 2-approximation.  Similarly, a $j$-approximation can be attained for any $j$, so \smaller{jPC-TEx} and \smaller{jPC-TMaxEx} are in APX for all $j$.

\begin{figure}[htbp]
	\begin{center}
		\includegraphics[width=12cm]{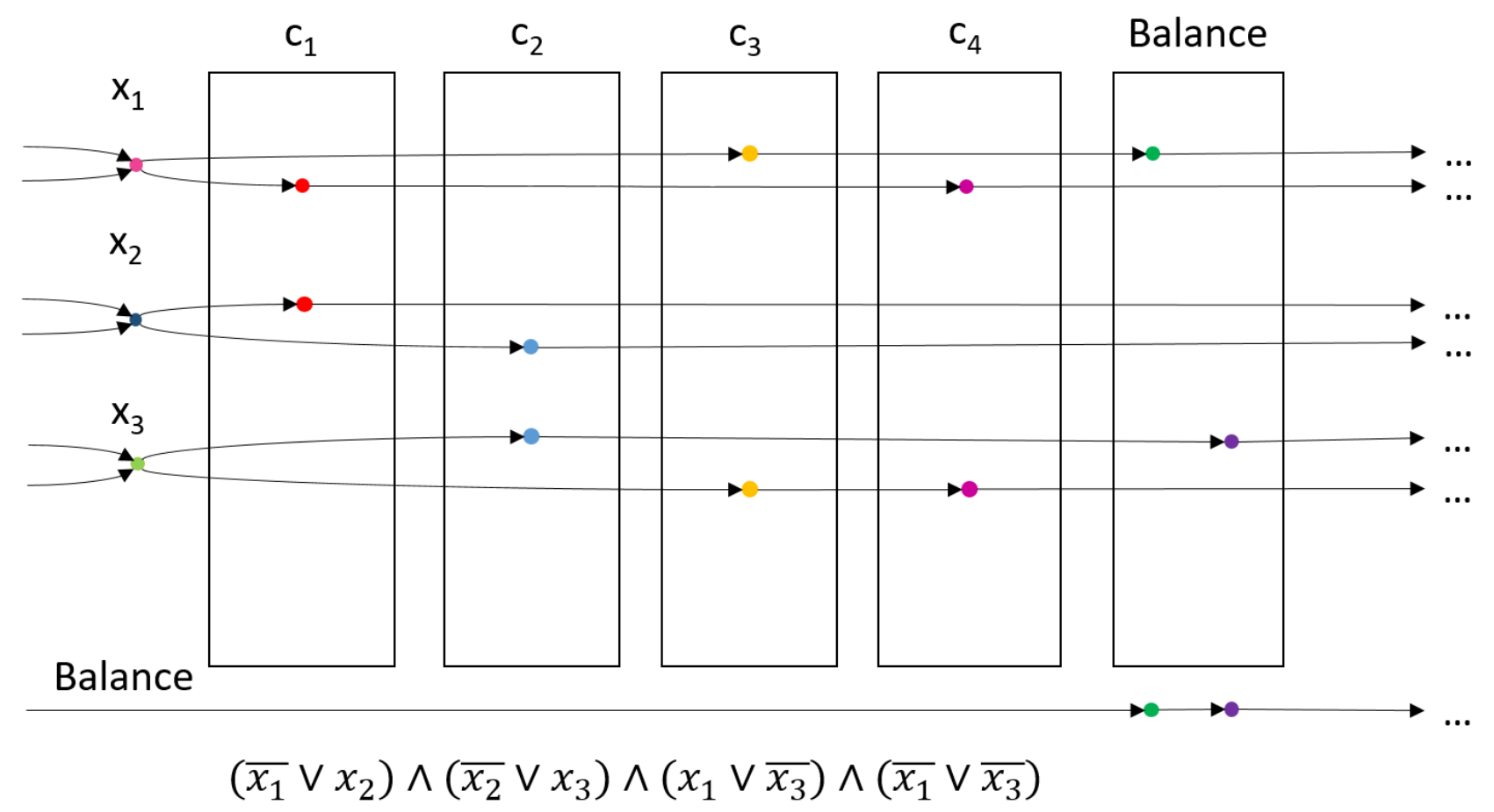}
	\end{center}
	\caption{MAX-2-SAT instance reduced to 2PC-TMaxEx.}
	\label{fig:2SAT_reduce_2PC}
\end{figure}

Consider an instance of the NP-hard problem \smaller{MAX-2-SAT}. Build a graph in polynomial time as described in Section \ref{se:rep_as_graph}, but instead of making all variable vertices the same color, assign a unique color to each variable vertex. This graph has at most two vertices of any given color, so it is an instance of \smaller{2PC-TEx}. Refer to Figure \ref{fig:2SAT_reduce_2PC}, but ignore the balance vertices for now. A solution of \smaller{2PC-TEx} will include as many colors as possible. If a clause's color is included in the solution, that corresponds to a satisfied clause, so \smaller{MAX-2-SAT} reduces to \smaller{2PC-TEx}, which means \smaller{2PC-TEx} is NP-hard, just as \smaller{MAX-2SAT} is NP-hard. 

Similarly, \smaller{2PC-TMaxEx} can be shown to be NP-hard with a reduction from \smaller{MAX-2-SAT}. As before, create a graph to represent the \smaller{MAX-2-SAT} instance, giving each variable vertex a unique color. Add balance vertices so that every cycle contains the same number of vertices.  Give each balance vertex its own unique color. Finally, add an additional cycle that includes a vertex for each of the balance vertex colors. Each balance vertex color appears only twice: once in a TRUE or FALSE cycle and once in the balance vertex cycle. Due to the balance vertex cycle, every balance vertex color is guaranteed to appear in the solution, so those colors do not affect any decisions regarding whether to take a TRUE cycle or a FALSE cycle. As a result, every combination of TRUE and FALSE cycles has the same number of vertices in the cycles, and a solution to \smaller{2PC-TMaxEx} will simply maximize the number of colors in the cycles. Maximizing the number of colors in the solution corresponds to maximizing the number of satisfied clauses. Thus, \smaller{2PC-TMaxEx} is also NP-hard. Figure \ref{fig:2SAT_reduce_2PC} depicts the reduction of an instance of \smaller{MAX-2-SAT} to \smaller{2PC-TMaxEx}.

Because \smaller{2PC-TEx} and \smaller{2PC-TMaxEx} are NP-hard, \smaller{jPC-TEx} and \smaller{jPC-TMaxEx} are also NP-hard for all $j>2$.

\begin{theorem}
	\smaller{jPC-TEx} and \smaller{jPC-TMaxEx} are NP-hard.
\end{theorem}

\section{\smaller{MAX-SIZE-TROPICAL-EXCHANGE} is NP-Hard}\label{se:MaxTROPICAL EXCHANGE}
For completeness, we consider \smaller{MAX-SIZE-TROPICAL-EXCHANGE-o (MaxTEx-o)}, where the primary criterion is maximizing the total number of colors in the set of vertex-disjoint cycles, and the secondary criterion is maximizing the number of vertices. This simply reverses the criteria of \smaller{TMaxEx-o}. It is easy to see that a solution to \smaller{MaxTEx-o} simultaneously solves \smaller{TEx-o}. One only needs to observe the number of colors in the solution to \smaller{MaxTEx-o}.  Thus, \smaller{MaxTEx-o} is NP-hard because \smaller{TEx-o} is NP-hard. Whether \smaller{MaxTEx-d} is in NP remains an open question. 

\section{Conclusion}\label{se:Conclusion}
In this paper, we have defined and analyzed problems that have practical application in the area of algorithmically arranged barter exchanges.  We have shown that \smaller{TROPICAL-EXCHANGE-d} and \smaller{TROPICAL-MAX-SIZE-EXCHANGE-d} are NP-complete and that \smaller{TROPICAL-EXCHANGE-o} and \smaller{TROPICAL-MAX-SIZE-EXCHANGE-o} are APX-hard.  We have also shown that \smaller{MAX-SIZE-TROPICAL-EXCHANGE-o} is NP-hard. When instances of \smaller{TROPICAL-EXCHANGE-o} and \smaller{TROPICAL-MAX-SIZE-EXCHANGE-o} are restricted to $j$ vertices per color (\smaller{jPC-TEx} and \smaller{jPC-TMaxEx}, respectively), the optimization problems remain NP-hard if $j > 1$, but are in APX.

\nocite{}
\bibliographystyle{abbrvnat}
\bibliography{highley_le-dmtcs}
\label{sec:biblio}

\end{document}